\documentclass[12pt]{article}
\usepackage{setspace}
\usepackage{fullpage}
\usepackage{amsbsy}
\usepackage{amsmath}
\usepackage{amsfonts}
\usepackage{amsthm}
\usepackage{amssymb}
\usepackage{algorithmic}
\usepackage{algorithm}
\usepackage{enumerate}
\usepackage{graphicx}
\usepackage{multirow}
\usepackage{natbib}
\setcitestyle{round}
\usepackage{subfigure}
\usepackage{listings}
\usepackage{xcolor}
\usepackage[maxfloats=40]{morefloats}
\usepackage{color}
\usepackage[font=footnotesize]{caption}
\theoremstyle{plain}

\theoremstyle{definition}
\newtheorem{experiment}{Experiment} 
\theoremstyle{remark}

\begin{document}

\title{\bf{Optimization and Testing in Linear Non-Gaussian Component Analysis}}

\author{Ze Jin, Benjamin B. Risk, David S. Matteson
\thanks{Research support from an NSF award (DMS-1455172), a Xerox PARC Faculty Research Award, and Cornell University Atkinson Center for a Sustainable Future (AVF-2017).}}

\date{\today}

\maketitle

\singlespacing
\begin{abstract}


Independent component analysis (ICA) decomposes multivariate data into mutually independent components (ICs).
The ICA model is subject to a constraint that at most one of these components is Gaussian,
which is required for model identifiability.
Linear non-Gaussian component analysis (LNGCA) generalizes the ICA model to a linear latent factor model with any number of both non-Gaussian components (signals)
and Gaussian components (noise), where observations are linear combinations of independent components.
Although the individual Gaussian components are not identifiable, the Gaussian subspace is identifiable.
We introduce an estimator along with its optimization approach
in which non-Gaussian 
and Gaussian components are estimated simultaneously,
maximizing the discrepancy of each non-Gaussian component from Gaussianity while minimizing the discrepancy of each Gaussian component from Gaussianity.
When the number of non-Gaussian components is unknown,
we develop a statistical test to determine it based on resampling and the discrepancy of estimated components.
Through a variety of simulation studies, we demonstrate the improvements of our estimator over competing estimators,
and we illustrate the effectiveness of the test to determine the number of non-Gaussian components.
Further, we apply our method to real data examples and demonstrate its practical value.

\end{abstract}
\doublespacing


\section{Introduction}\label{intro}

Independent component analysis (ICA) finds a representation of multivariate data based on mutually independent components (ICs).
As an unsupervised learning method, ICA has been developed for applications including
blind source separation, feature extraction, brain imaging, and many others.
\citet{hyvarinen2004independent} provided an overview of ICA approaches for measuring the non-Gaussianity and estimating the ICs.
%

Let $Y = (Y_1, \dots, Y_p)^T \in \mathbb{R}^p$ be a random vector of observations.
Assume that $Y$ has a nonsingular, continuous distribution $F_Y$, with $\textrm{E}(Y_j) = 0$
and $\textrm{Var}(Y_j) < \infty$, $j = 1, \dots, p$.
Let $X = (X_1, \dots, X_p)^T \in \mathbb{R}^p$ be a random vector of latent components.
Without loss of generality, $X$ is assumed to be standardized such that
$\textrm{E}(X_j) = 0$ and $\textrm{Var}(X_j) = 1$, $j = 1, \dots, p$.
A static linear latent factor model to estimate the components $X$ from the observations $Y$ is given by
\begin{eqnarray*}\label{ica1}
\begin{aligned}
Y &= AX,\\
X &= A^{-1}Y \triangleq BY
\end{aligned}
\end{eqnarray*}
where $A \in \mathbb{R}^{p \times p}$ is a constant, nonsingular mixing matrix,
and $B \in \mathbb{R}^{p \times p}$ is a constant, nonsingular unmixing matrix.

Pre-whitened random variables are uncorrelated and thus easier to work with from both practical and theoretical perspectives.
Let $\Sigma_Y = \textrm{Cov}(Y)$ be the covariance matrix of $Y$,
and $H = \Sigma_Y^{-1/2}$ be an uncorrelating matrix.
Let $Z = HY = (Z_1, \dots, Z_p)^T \in \mathbb{R}^p$ be a random vector of uncorrelated observations,
such that $\Sigma_Z = \textrm{Cov}(Z) = I_p$, the $p \times p$ identity matrix.
The ICA model further assumes that the components $X_1, \dots, X_p$ are mutually independent,
in which the number of Gaussian components is at most one.
Then the relationship between $X$ and $Z$ in the ICA model is
\begin{eqnarray}\label{ica3}
\begin{aligned}
X &= A^{-1}Y = A^{-1}H^{-1}Z \triangleq WZ = M^TZ,\\
Z &= W^{-1}X = HAX \triangleq MX = W^TX
\end{aligned}
\end{eqnarray}
where $W = A^{-1}H^{-1} \in \mathbb{R}^{p \times p}$ is a constant, nonsingular unmixing matrix,
and $M = HA \in \mathbb{R}^{p \times p}$ is a constant, nonsingular mixing matrix.
Given that $Z$ are uncorrelated observations, $W$ is an orthogonal matrix,
and $M$ is an orthogonal matrix as well. Thus, we have $W = M^{-1} = M^T$ and $M = W^{-1} = W^T$.

Many methods have been proposed for estimating the ICA model in the literature,
including the fourth-moment diagonalization of FOBI \citep{cardoso1989source}
and JADE \citep{cardoso1993blind},
the information criterion of Infomax \citep{bell1995information},
maximizing negentropy in FastICA \citep{hyvarinen1997fast},
the maximum likelihood principle of ProDenICA \citep{hastie2003independent},
and the mutual dependence measure of dCovICA \citep{matteson2017independent}
and MDMICA \citep{jin2017independent}.
Most of them use optimization to obtain the components such that
they have maximal non-Gaussianity under the constraint that they are uncorrelated.
The goal is to use $Z$ to estimate both $W$ and $X$, by maximizing the non-Gaussianity of the components in $X$,
according to a particular measure of non-Gaussianity.

To overcome the limit of the ICA model that at most one Gaussian component exists, the NGCA (non-Gaussian component analysis) model was proposed \citet{blanchard2006non}.
Beginning with (\ref{ica3}), the components $X \in \mathbb{R}^p$ are decomposed into signals $S \in \mathbb{R}^q$ and noise $N \in \mathbb{R}^{p-q}$,
and $M$ into $M_S$ and $M_N$, and $W$ into $W_S$ and $W_N$ correspondingly.
The components in $S$ are assumed to be non-Gaussian, while the components in $N$ are assumed to be Gaussian.
The NGCA model further assumes that the non-Gaussian components $S$ are independent of the Gaussian components $N$,
the components in $N$ are mutually independent and thus are multivariate normal,
although the components in $S$ may remain mutually dependent.
Then the relationship between $X$ and $Z$ in the NGCA model is
\begin{eqnarray}\label{ngca}
\begin{aligned}
\begin{bmatrix}
                          S \\
                          N \\
                        \end{bmatrix} &= X = WZ =  \begin{bmatrix}
                          W_S Z \\
                          W_N Z \\
                        \end{bmatrix}, \\
Z &= MX = \begin{bmatrix}
           M_S & M_N \\
         \end{bmatrix} \begin{bmatrix}
                          S \\
                          N \\
                        \end{bmatrix} = M_S S + M_N N
\end{aligned}
\end{eqnarray}
where $M_S \in \mathbb{R}^{p \times q}$ has rank $q$, $M_N \in \mathbb{R}^{p \times (p-q)}$ has rank $p-q$,
$W_S \in \mathbb{R}^{q \times p}$ has rank $q$, and $W_N \in \mathbb{R}^{(p-q) \times p}$ has rank $p-q$.
The goal is to estimate the non-Gaussian subspace spanned by
the rows in $W_S$ corresponding to $S$,
as the Gaussian subspace corresponding to $N$ is uninteresting.
\cite{kawanabe2007new} developed an improved algorithm based on radial kernel functions.
\citet{theis2011uniqueness} proved a necessary and sufficient condition for the uniqueness
of the non-Gaussian subspace from projection methods.
\citet{bean2014non} developed theory for an approach based on characteristic functions.
\citet{sasaki2016non} introduced a least-squares NGCA (LSNGCA) algorithm based on
least-squares estimation of log-density gradients and eigenvalue decomposition,
and \citet{shiino2016whitening} proposed a whitening-free variant of LSNGCA.
\citet{nordhausen2017asymptotic} developed asymptotic and bootstrap tests for the dimension
of non-Gaussian subspace based on the FOBI method.

To incorporate nice characteristics from both the ICA model and NGCA model,
we consider the LNGCA (linear non-Gaussian component analysis) model proposed in \citet{risk2016linear}
as a special case of the NGCA model, which is the same as the the NGICA model in \citet{virta2016projection}.
In the form of (\ref{ngca}),
the LNGCA model further assumes that the components $X_1, \dots, X_p$ are mutually independent,
and allows any number of both non-Gaussian components and Gaussian components among them.
Similarly, we have $W = M^{-1} = M^T$ and $M = W^{-1} = W^T$.
Then the relationship between $X$ and $Z$ in the LNGCA model is
\begin{eqnarray*}\label{LNGCA}
\begin{aligned}
&&\begin{bmatrix}
                          S \\
                          N \\
                        \end{bmatrix} = X = WZ
                        = \begin{bmatrix}
                          W_S Z\\
                          W_N Z\\
                        \end{bmatrix}
                        = M^{T}Z = \begin{bmatrix}
                          M_S^T Z \\
                          M_N^T Z \\
                        \end{bmatrix}, \\
&& Z = MX = \begin{bmatrix}
           M_S & M_N \\
         \end{bmatrix} \begin{bmatrix}
                          S \\
                          N \\
                        \end{bmatrix} = M_S S + M_N N
\end{aligned}
\end{eqnarray*}
where $M_S \in \mathbb{R}^{p \times q}$ has rank $q$, $M_N \in \mathbb{R}^{p \times (p-q)}$ has rank $p-q$,
$W_S \in \mathbb{R}^{q \times p}$ has rank $q$, and $W_N \in \mathbb{R}^{(p-q) \times p}$ has rank $p-q$.
\citet{risk2016linear} presented a parametric LNGCA using the logistic density and
a semi-parametric LNGCA using tilted Gaussians with cubic B-splines to estimate this model.
\citet{virta2016projection} used projection pursuit to extract the non-Gaussian components
and separate the corresponding signal and noise subspaces
where the projection index is a convex combination of squared third and fourth cumulants.


In this paper, we study the LNGCA model by taking advantage of its flexibility in the number of Gaussian components,
and mutual independence assumption between all components.
With pre-whitening, the Gaussian contribution to the model likelihood is invariant to linear transformations that preserve unit variance, as shown in \cite{risk2016linear}. Thus, an alternative framework is necessary in order to leverage the information in the Gaussian subspace. This motivates our novel objective function, which estimates the unmixing matrix $W$ by maximizing the discrepancy from Gaussianity for the non-Gaussian components and \emph{minimizing} the discrepancy for the Gaussian components, thereby explicitly estimating the Gaussian subspace to improve upon constrained maximum likelihood approaches.
The rest of this paper is organized as follows.
In Section \ref{disc}, we introduce the discrepancy functions to measure the distance from Gaussianity.
In Section \ref{opt}, we propose a framework of LNGCA estimation given the number of non-Gaussian components $q$.
In Section \ref{test},
we introduce a sequence of statistical tests to determine the number of non-Gaussian components $q$ when it is unknown.
We present the simulation results in Section \ref{sim}, followed by real
data examples in Section \ref{data}. Finally, Section \ref{con} is the summary of our work.

The following notations will be used throughout this paper.
Let $\mathcal{O}_{a \times b}$ denote the set of $a \times b$ matrices whose columns are orthonormal.
Let $\mathcal{P}^{\pm}_{a \times a}$ denote the set of $a \times a$ signed permutation matrices.
Let $||U||_\textrm{F} = \sqrt{\sum_{i,j}U_{ij}^2}$ denote the Frobenius norm of $U \in \mathbb{R}^{a \times b}$.

\section{Discrepancy}\label{disc}

\subsection{Population Discrepancy Measures}\label{disc1}

In order to find the best estimate for the LNGCA model, we need a criterion to measure the discrepancy between $X$ and its underlying assumption,
i.e., $S$ should be far from Gaussianity and $N$ should be close to Gaussianity.
Specifically, we choose a general class of functions $\mathcal{D}$ that measure the discrepancy $D$ between each component $X_j$ and Gaussianity.

\citet{hastie2003independent} proposed the expected log-likelihood tilt function
to measure the discrepancy from Gaussianity in the estimation of the ICA model.  
Suppose the density of $X_j$ is $f_j$, $j = 1, \dots, p$,
and each of the densities $f_j$ is represented by an exponentially tilted Gaussian density
\begin{equation*}\label{den1}
f_j(x_j) = \phi(x_j)e^{g_j(x_j)}
\end{equation*}
where $\phi$ is the standard univariate Gaussian density, and $g_j$ is a smooth function.
The log-tilt function $g_j$ represents departures from Gaussianity,
and the expected log-likelihood ratio between
$f_j$ and the Gaussian density is
\begin{equation*}\label{GPois1}
\textrm{GPois}(X_j) = \textrm{E}[g_j(X_j)].
\end{equation*}

\citet{virta2015joint, virta2016projection} proposed the use of the Jarque-Bera (JB) test statistic \citep{jarque1987test}
\begin{equation*}\label{JB1}
\textrm{JB}({X}_j) = \textrm{Skew}({X}_j) + \frac{1}{4}\textrm{Kurt}({X}_j)
\end{equation*}
to measure the discrepancy from Gaussianity in the estimation of ICA and LNGCA models,
where
\begin{eqnarray*}\label{skewkurt1}
\begin{aligned}
\textrm{Skew}({X}_j) &= \left(
                                \textrm{E} [{X}_{j}^3]
                              \right)^2,\\
\textrm{Kurt}({X}_j) &= \left(
                                \textrm{E} [{X}_{j}^4] - 3
                              \right)^2
\end{aligned}
\end{eqnarray*}
are squared skewness and squared excess kurtosis.
In fact, \citet{virta2015joint, virta2016projection} studied a linear combination of Skew and Kurt,
i.e., $\alpha \textrm{Skew} + (1 - \alpha) \textrm{Kurt}$,
and advised the choice of $\alpha = 0.8$, which corresponds to JB.
This takes deviation of both skewness and kurtosis into account,
while Skew and Kurt are valid discrepancy functions as well.
Notice that JB$(X_j)$, Skew$(X_j)$, and Kurt$(X_j)$ are simplified due to standardized ${X}_j$.

\subsection{Empirical Discrepancy Measures}\label{disc2}

Let $\mathbf{Y} = \{Y^i = (Y_1^i, \dots, Y_p^i): i = 1, \dots, n \} \in \mathbb{R}^{n \times p}$ be an i.i.d.\ sample of observations from $F_Y$,
and let $\mathbf{Y}_j = \{Y^i_j: i = 1, \dots, n \} \in \mathbb{R}^p$ be the corresponding i.i.d.\ sample of observations from $F_{Y_j}$,
$j = 1, \dots, p$, such that $\mathbf{Y} = \{\mathbf{Y}_1, \dots, \mathbf{Y}_p\}$.
Let $\widehat{\Sigma}_\mathbf{Y}$ be the sample covariance matrix of $\mathbf{Y}$,
and $\widehat{H} = \widehat{\Sigma}_\mathbf{Y}^{-1/2}$ be the estimated uncorrelating matrix.
Although the covariance $\Sigma_Y$ is unknown in practice,
the sample covariance $\widehat{\Sigma}_\mathbf{Y}$ is a consistent estimate under the finite second-moment assumption.
Let ${\widehat{\mathbf{Z}}} = \mathbf{Y} \widehat{H}^T \in \mathbb{R}^{n \times d}$ be the estimated uncorrelated observations,
such that $\widehat{\Sigma}_{{\widehat{\mathbf{Z}}}} = I_d$,
and ${\Sigma}_{{\widehat{\mathbf{Z}}}} \overset{a.s.}{\longrightarrow} I_d$ as $n \rightarrow \infty$.

To simplify notation, we assume that ${\mathbf{Z}}$, an uncorrelated i.i.d.\ sample is given with mean zero and unit variance.
Let $\mathbf{X} = \{X^i = (X_1^i, \dots, X_p^i): i = 1, \dots, n \} =
[\mathbf{S}, \mathbf{N}] = \mathbf{Z}W^T \in \mathbb{R}^{n \times p}$ be the sample of $X$,
where $\mathbf{S} \in \mathbb{R}^{n \times q}$ and $\mathbf{N} \in \mathbb{R}^{n \times (p-q)}$,
and let $\mathbf{X}_j = \{X^i_j: i = 1, \dots, n \} \in \mathbb{R}^n$ be the sample of $X_j$,
i.e., the $j$th column in $\mathbf{X}$.
Similarly, we can define
$\mathbf{S}_j, \mathbf{N}_j \in \mathbb{R}^n$.
Notice that $\mathbf{X}_j, \mathbf{S}_j, \mathbf{N}_j$ has sample mean 0 and sample variance 1.

We obtain the empirical discrepancy $\widehat{D}$ by replacing expectations by sample averages.
The empirical GPois is given by
\begin{equation*}\label{GPois2}
\textrm{GPois}(\mathbf{X}_j) = \frac{1}{n}\sum_{i=1}^n \widehat{g}_j(X_j^i)
\end{equation*}
where $\widehat{g}_j$ is estimated by maximum penalized likelihood, maximizing the criterion
\begin{equation*}\label{mle2}
\sum_{j=1}^p \left\{  \frac{1}{n} \sum_{i=1}^n \left[\log \phi(X_{j}^i) + \widehat{g}_j(X_{j}^i)\right] - \lambda_j \int \widehat{g}_j''^2(x)dx \right\}
\end{equation*}
subject to
\begin{equation*}\label{int2}
\int \phi(s)e^{\widehat{g}_j(x)} \, dx = 1
\end{equation*}
where $\widehat{g}_j$ is estimated by a smoothing spline, and $\lambda_j$ is selected by
controlling the degrees of freedom of the smoothing spline, which is 6 by default
in the R package \texttt{ProDenICA} \citep{hastie2010prodenica}.

The empirical JB is given by
\begin{equation*}\label{JB2}
\textrm{JB}(\mathbf{X}_j) = \textrm{Skew}(\mathbf{X}_j) + \frac{1}{4}\textrm{Kurt}(\mathbf{X}_j)
\end{equation*}
where
\begin{eqnarray*}\label{skewkurt2}
\begin{aligned}
\textrm{Skew}(\mathbf{X}_j) &= \left(
                                \frac{1}{n} \sum_{k=1}^n (X_j^k)^3
                              \right)^2,\\
\textrm{Kurt}(\mathbf{X}_j) &= \left(
                                \frac{1}{n} \sum_{k=1}^n (X_j^k)^4 - 3
                              \right)^2
\end{aligned}
\end{eqnarray*}
are the empirical Skew and empirical Kurt.
We will see that JB (joint use of skewness and kurtosis) performs much better than
either Skew (use of skewness only) or Kurt (use of kurtosis only) alone in the simulations of Section \ref{sim},
which was shown in \citet{virta2016projection} as well.

\section{Optimization Strategy}\label{opt}
Using $\widehat{D}$ to measure the difference between $\mathbf{X}_j$ and Gaussianity,
we seek an optimal $W$ such that $\mathbf{X}$ is most likely to fit the underlying model with
independent components.

For the ICA model, a classical ICA estimator to estimate $W$ in
%
FastICA \citep{hyvarinen1997fast} and ProDenICA \citep{hastie2003independent} is defined by
\begin{equation*}\label{est1}
\widehat{W}^\ast = \underset{W \in \mathcal{O}_{p \times p}}{\arg\max} \sum_{j=1}^p \widehat{D}(\mathbf{X}_j).
\end{equation*}

We can naturally extend the ICA estimator to an LNGCA estimator given $q$ as
\begin{equation}\label{est2}
\widehat{W}_{S}^{\textrm{max}} = \underset{W \in \mathcal{O}_{p \times q}}{\arg\max} \sum_{j:X_j \in S} \widehat{D}(\mathbf{X}_j)
= \underset{W \in \mathcal{O}_{p \times q}}{\arg\max} \sum_{j=1}^q \widehat{D}(\mathbf{S}_j)
\end{equation}
which is named the max estimator,
as we maximize the discrepancy between non-Gaussian components and Gaussianity.
The algorithm for the max estimator is described in Alg.\ \ref{alg1},
where the fixed point algorithm is elaborated in \citet{hastie2003independent}.
The objective function used in Spline-LCA from \citet{risk2016linear} is the same as the max estimator
when $f$ is GPois, but the optimization differs, which will be explored in Section \ref{sim}.

Given the estimated unmixing matrix $\widehat{W}^{\textrm{max}}_{S}$,
the estimated non-Gaussian components are $\widehat{\mathbf{S}} = \mathbf{Z}(\widehat{W}^{\textrm{max}}_{S})^T$.
\begin{algorithm}
\caption{LNGCA algorithm for the max estimator}\label{alg1}
\begin{algorithmic}
\STATE 1. Initialize $W_{p \times q}$.
\STATE 2. Alternate until convergence of $W$, using the Frobenius norm.
\STATE \quad (a) Given $W$, estimate the discrepancy $\widehat{D}(\mathbf{S}_j)$ of component $\mathbf{S}_j$ for each $j$.
\STATE \quad (b) Given $\widehat{D}(\mathbf{S}_j)$, $j = 1, \dots, q$, perform one step of the fixed point algorithm towards finding the optimal $W$.
\end{algorithmic}
\end{algorithm}

Since any rotation of a Gaussian distribution will lead to the same Gaussian distribution,
the Gaussian components $N$ are not identifiable.
However, we can benefit from estimating the Gaussian subspace for the LNGCA model, since the column space of $W_N$ is identifiable.
Taking $N$ into account by optimizing $S$ and $N$ simultaneously in the objective function,
we expect to recognize the Gaussian subspace, which helps shape the non-Gaussian subspace
because the non-Gaussian subspace is the complement of the Gaussian subspace.
Motivated by this optimization idea, we propose a new LNGCA estimator given $q$ as
\begin{equation}\label{est3}
\widehat{W}^{\textrm{max-min}} = \underset{{W \in \mathcal{O}_{p \times p}}}{\arg\max} \left[\sum_{j:X_j \in S} \widehat{D}(\mathbf{X}_j) - \sum_{j:X_j \in N} \widehat{D}(\mathbf{X}_j)\right] =
\underset{{W \in \mathcal{O}_{p \times p}}}{\arg\max} \left[\sum_{j=1}^q \widehat{D}(\mathbf{S}_j) - \sum_{j=1}^{p-q} \widehat{D}(\mathbf{N}_j)\right]
\end{equation}
which is named the max-min estimator for the LNGCA model,
as we maximize the discrepancy between non-Gaussian components and Gaussianity,
and minimize the discrepancy between Gaussian components and Gaussianity simultaneously.
The algorithm for the max-min estimator is described in Alg.\ \ref{alg2},
where the fixed point algorithm is elaborated in \citet{hastie2003independent}.
We will see that the max-min estimator (joint optimization of $S$ and $N$) performs much better than
the max estimator (optimization of $S$ only) in the simulations of Section \ref{sim}.
\begin{algorithm}
\caption{LNGCA algorithm for the max-min estimator}\label{alg2}
\begin{algorithmic}
\STATE 1. Initialize $W_{p \times p}$.
\STATE 2. Alternate until convergence of $W$, using the Frobenius metric.
\STATE \quad (a) Given $W$, estimate the discrepancy $\widehat{D}(\mathbf{X}_j)$
of component $\mathbf{X}_j$ for each $j$.
\STATE \quad (b) Sort components by discrepancy $\widehat{D}(\mathbf{X}_j)$ in decreasing order.
\STATE \quad (c) Flip the sign of discrepancy $\widehat{D}(\mathbf{X}_j)$ of the last $p - q$ components.
\STATE \quad (d) Given $\widehat{D}(\mathbf{X}_j)$, $j = 1, \dots, p$, perform one step of the fixed point algorithm towards finding the optimal $W$.
\STATE 3. Sort components by discrepancy $\widehat{D}(\mathbf{X}_j)$ in decreasing order.
\end{algorithmic}
\end{algorithm}

Given the estimated unmixing matrix $\widehat{W}^{\textrm{max-min}}$,
the estimated non-Gaussian and Gaussian components are $\widehat{\mathbf{X}} = \mathbf{Z}(\widehat{W}^{\textrm{max-min}})^T$.
However, it is not clear which component in $\widehat{\mathbf{X}}$ belongs to $\widehat{\mathbf{S}}$ or $\widehat{\mathbf{N}}$,
since $\widehat{\mathbf{S}}$ and $\widehat{\mathbf{N}}$ are obtained together instead of $\widehat{\mathbf{S}}$ only.
The solution is to sort the independent components $X_1, \dots, X_p$ by discrepancy value $D(X_i)$ in decreasing order,
and obtain the ordered independent components $X_{(1)}, \dots, X_{(p)}$.
Given that there are $q$ non-Gaussian components, it is natural to take
$S = (X_{(1)}, \dots, X_{(q)})^T$ and $N = (X_{(q+1)}, \dots, X_{(p)})^T$ based on the discrepancy function measuring non-Gaussianity.
As the $q$ non-Gaussian components in $S$
have the $q$-largest discrepancy values $D$ among $X_1, \dots, X_p$,
the estimated non-Gaussian components in $\widehat{\mathbf{S}}$ are expected to have the $q$-largest empirical discrepancy values $\widehat{D}$ among
$\mathbf{\widehat{X}}_1, \dots, \mathbf{\widehat{X}}_p$.

Nevertheless, we cannot sort $\mathbf{X}$ by the empirical discrepancy to determine
which component in $\mathbf{X}$ belongs to $\mathbf{S}$ or $\mathbf{N}$ at the beginning,
and then stick to the order throughout the iterative algorithm
and conclude which component in $\widehat{\mathbf{X}}$ belongs to $\widehat{\mathbf{S}}$ or $\widehat{\mathbf{N}}$ in the end,
since the optimization does depend on the initialization,
and the order of components may change after each iteration.
Instead, we repeatedly sort $\mathbf{X}$ by empirical discrepancy value
and adaptively determine the components in $\mathbf{S}$ and $\mathbf{N}$ at the end of each iteration in Alg \ref{alg2}.
Finally, when the algorithm converges, we sort the estimated components $\mathbf{\widehat{X}}_1, \dots, \mathbf{\widehat{X}}_p$
by empirical discrepancy value,
and obtain the ordered estimated components $\mathbf{\widehat{X}}_{(1)}, \dots, \mathbf{\widehat{X}}_{(p)}$.
Then we take $\widehat{\mathbf{S}} = [\mathbf{\widehat{X}}_{(1)}, \dots, \mathbf{\widehat{X}}_{(q)}]$,
and $\widehat{\mathbf{N}} = [\mathbf{\widehat{X}}_{(q+1)}, \dots, \mathbf{\widehat{X}}_{(p)}]$.
Accordingly, we decompose $\widehat{W}$ into $\widehat{W}_S$ and $\widehat{W}_N$,
and $\widehat{M} = \widehat{W}^T$ into $\widehat{M}_S$ and $\widehat{M}_N$.

\section{Testing and Subspace Estimation}\label{test}

In practice, the number of non-Gaussian components $q$ is unknown.
Following the convention of ordered components with respect to non-Gaussianity,
we introduce a sequence of statistical tests to decide $q$.
The main idea is that, for any $j' < j$,
$X_{(j')}$ is more likely to be non-Gaussian than $X_{(j)}$ in terms of discrepancy value $D$.
If there are $k$ non-Gaussian independent components,
then $X_{(1)}, \dots, X_{(k)}$ are non-Gaussian, and $X_{(k+1)}, \dots, X_{(p)}$ are Gaussian.

Based on this heuristic, we propose a sequence of hypotheses for searching $q$ as
\begin{eqnarray*}\label{H0}
\begin{aligned}
H_0^{(k)}&: X_{(1)}, \dots, X_{(k-1)} \textrm{ are non-Gaussian and } X_{(k)}, \dots, X_{(p)} \textrm{ are Gaussian}, \\
H_A^{(k)}&: X_{(1)}, \dots, X_{(k)} \textrm{ are non-Gaussian}
\end{aligned}
\end{eqnarray*}
which is equivalent to testing whether there are exactly $k-1$ non-Gaussian components
or at least $k$ non-Gaussian components.

Under $H_0^{(k)}$, we first run the optimization from $\mathbf{X} = \mathbf{Z}W^T$
using the max-min estimator with $q = k-1$, in which we estimate $\widehat{W}_{}$ and
$\widehat{\mathbf{X}} = [\mathbf{\widehat{X}}_{(1)}, \dots, \mathbf{\widehat{X}}_{(p)}]$ from the sample data $\mathbf{Z}$.
One thing worth mentioning is that $\widehat{\mathbf{X}}$ depends on $k$
as the optimization depends on $k$, although we suppress the notation here.

Next we repeat the following resampling procedure for $B$ times:
during the $b$th time,
we randomly generate independent Gaussian $\mathbf{G}^{(b)} = [\mathbf{G}^{(b)}_1, \dots, \mathbf{G}^{(b)}_{p-k+1}]$
with the same number of observations as $\mathbf{Z}$, and construct pseudo components
$\mathbf{X}^{(b)} = [\mathbf{\widehat{X}}_{(1)}, \dots, \mathbf{\widehat{X}}_{(k-1)}, \mathbf{G}^{(b)}]$.
Based on the estimated unmixing matrix $\widehat{W}_{}$,
we use the estimated mixing matrix $\widehat{M} = \widehat{W}_{}^T$
to construct pseudo observations $\mathbf{Z}^{(b)} = \mathbf{X}^{(b)}\widehat{M}^T$.
Then we run the optimization from $\mathbf{X}^{(b)} = \mathbf{Z}^{(b)}{W^T_{}}$
using the max-min estimator with $q=k-1$, and we estimate $\widehat{W}^{(b)}$ and
$\widehat{\mathbf{X}}^{(b)} = [\mathbf{\widehat{X}}_{(1)}^{(b)}, \dots, \mathbf{\widehat{X}}^{(b)}_{(p)}]$
from the pseudo data $\mathbf{Z}^{(b)}$.

At last, we calculate an approximate p-value by
comparing $\widehat{D}(\widehat{\mathbf{X}}_{(k)})$ to $\widehat{D}(\widehat{\mathbf{X}}^{(b)}_{(k)})$,
or $\sum_{j=1}^k \widehat{D}(\widehat{\mathbf{X}}_{(j)})$ to $\sum_{j=1}^k \widehat{D}(\widehat{\mathbf{X}}^{(b)}_{(j)})$ as
\begin{eqnarray}\label{pval}
\begin{aligned}
\widehat{p}_{\textrm{curr}} &= \frac{\#\left\{\widehat{D}(\widehat{\mathbf{X}}_{(k)}) \leq \widehat{D}(\widehat{\mathbf{X}}^{(b)}_{(k)})\right\}}{B},\\
\widehat{p}_{\textrm{cumu}} &= \frac{\#\left\{\sum_{j=1}^k \widehat{D}(\widehat{\mathbf{X}}_{(j)}) \leq \sum_{j=1}^k \widehat{D}(\widehat{\mathbf{X}}^{(b)}_{(j)})\right\}}{B}
\end{aligned}
\end{eqnarray}
which we name the current method and the cumulative method respectively.

Our test shares the resampling technique with \citet{nordhausen2017asymptotic}.
However, there are two major differences. On the one hand,
our test does not need to bootstrap on $\mathbf{X}$, and thus saves remarkable computational cost,
and we will show that it accurately estimates the number of components.
On the other hand, our test is more flexible on the test statistic,
as it does not need to match what is used in the objective function in the optimization.
The algorithm for our sequential test is summarized in Alg.\ \ref{alg3} below.
\begin{algorithm}
\caption{The algorithm for the sequential test $H_0^{(k)}$}\label{alg3}
\begin{algorithmic}
\STATE 1. Estimate $\widehat{W}$ from $\mathbf{X} = \mathbf{Z}W^T$ using the max-min estimator with $q = k - 1$.
\STATE 2. Estimate $\widehat{\mathbf{X}} = \mathbf{Z}\widehat{W}^T = [\mathbf{\widehat{X}}_{(1)}, \dots, \mathbf{\widehat{X}}_{(p)}]$.
\STATE 3. Repeat the procedure for $B$ times:
\STATE \quad (a) Generate independent Gaussian $\mathbf{G}^{(b)} = [\mathbf{G}^{(b)}_1, \dots, \mathbf{G}^{(b)}_{p-k+1}]$.
\STATE \quad (b) Construct $\mathbf{X}^{(b)} = [\mathbf{\widehat{X}}_{(1)}, \dots, \mathbf{\widehat{X}}_{(k-1)}, \mathbf{G}^{(b)}]$.
\STATE \quad (c) Construct $\mathbf{Z}^{(b)} = \mathbf{X}^{(b)}\widehat{M}^T = \mathbf{X}^{(b)}\widehat{W}$.
\STATE \quad (d) Estimate $\widehat{W}^{(b)}$ from $\mathbf{X}^{(b)} = \mathbf{Z}^{(b)}{W^T_{}}$ using the max-min estimator with $q = k - 1$.
\STATE \quad (e) Estimate $\widehat{\mathbf{X}}^{(b)} = \mathbf{Z}^{(b)}({\widehat{W}^{{b}}})^T = [\mathbf{\widehat{X}}_{(1)}^{(b)}, \dots, \mathbf{\widehat{X}}^{(b)}_{(p)}]$.
\STATE 3. Calculate p-value using the current or cumulative method in (\ref{pval}).
\end{algorithmic}
\end{algorithm}


The proposed procedure involves a sequence of tests, but the number of tests can be dramatically reduced by using a binary search.
This approach quickly narrows in on the selected $q$
because we focus on the boundary that the p-value crosses a specific significance level.
As we expect no more than $\lceil\log_2 p\rceil$ tests, it makes sense to apply the Bonferroni correction.
Note that even for fairly large $p$, the number of tests remains reasonable, e.g., $p = 10,000$ implies fewer than fourteen tests.
Multiple testing in this setting of sequential testing may become more problematic as the dimension or search space grows,
though the sequential searching works well in the simulations of Section \ref{sim}.
Issues with multiple testing is an important direction for future research.

\section{Simulation Study}\label{sim}

\subsection{Sub- and Super-Gaussian Densities}
In this section, we evaluate the performance of the max-min estimator
by performing simulations similar to \citet{matteson2017independent} for the LNGCA model,
and compare it to that of the max estimator
using several discrepancy functions including Skew, Kurt, JB, GPois, and Spline.
Moreover, we elaborate on the implementation and performance measure of the LNGCA model.

\begin{figure}[ht]
\begin{center}
\centerline{\includegraphics[width=\columnwidth]{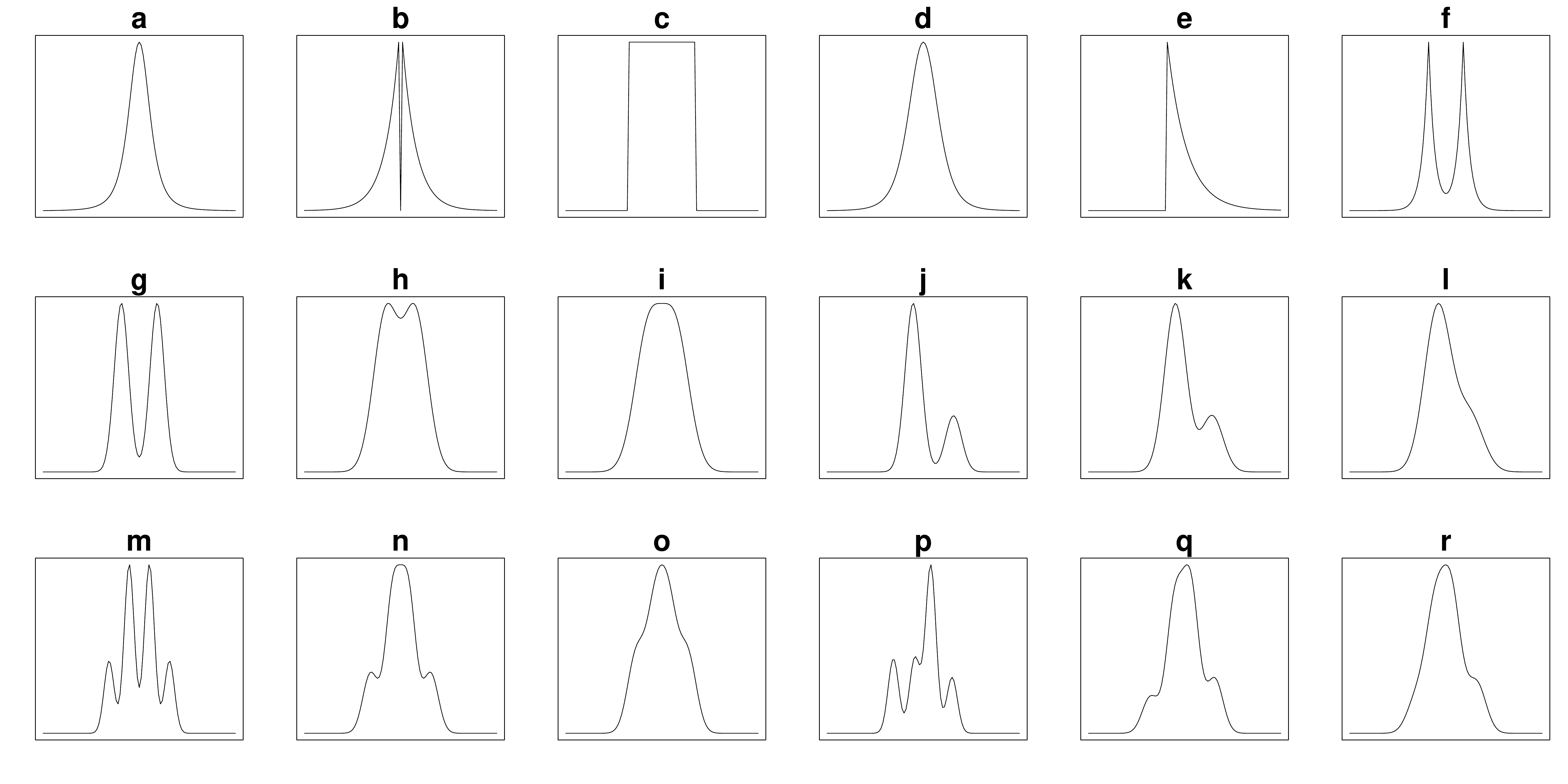}}
\caption{Density plots of the 18 distributions from \texttt{rjordan} in the R package \texttt{ProDenICA}.}
\label{fig1}
\end{center}
\end{figure}

We generate the non-Gaussian independent components $\mathbf{S} \in \mathbb{R}^{n \times q}$
from 18 distributions using \texttt{rjordan}
in the R package \texttt{ProDenICA} \citep{hastie2010prodenica} with sample size $n$ and dimension $q$.
See Figure \ref{fig1} for the density functions of the 18 distributions.
We also generate the Gaussian independent components $\mathbf{N} \in \mathbb{R}^{n \times (p-q)}$
with sample size $n$ and dimension $p - q$.
Then $\mathbf{X} = [\mathbf{S}, \mathbf{N}]$ are the underlying components of interest.
We simulate a mixing matrix $A \in \mathbb{R}^{p\times p}$ with condition number
between 1 and 2 using \texttt{mixmat} in the R package \texttt{ProDenICA} \citep{hastie2010prodenica}
and obtain the observations $\mathbf{Y} = \mathbf{X}A^T$,
which are centered by their sample mean, then pre-whitened by their sample covariance to obtain uncorrelated observations
$\mathbf{Z} = \mathbf{Y}\widehat{H}^T$.
Finally, we estimate $\widehat{W}_S$ and $\widehat{M}_S = \widehat{W}_S^T$
based on $\mathbf{Z}$ via the max estimator or the max-min estimator.
Therefore, $\mathbf{Z} = \mathbf{X} A^T\widehat{H}^T = \mathbf{X} (\widehat{H}A)^T$,
and we evaluate the estimation by comparing the estimated unmixing matrix
$\widehat{W}$ to the ground truth
$W^0 = (\widehat{H}A)^{-1} = A^{-1}\widehat{H}^{-1} = B\widehat{H}^{-1}$ with respect to $S$,
i.e., comparing $\widehat{W}_S$ to $W_S^0$ where
$W_S^0 = B_S\widehat{H}^{-1}$.


The optimization problem associated with the max estimator in (\ref{est2}) and the max-min estimator in (\ref{est3}) is non-convex,
which requires the initialization step and is sensitive to the initial point.
\cite{risk2014evaluation} demonstrated strong sensitivity to the initialization matrix in various ICA algorithms for the eighteen distributions considered in the experiments below.
To mitigate the presence of local maximum, we explore two options, one with a single initial point,
and another with multiple initial points, where each initial point is generated by orthogonalizing matrices with random Gaussian elements.
We suggest that the number of multiple initial points $m$ should grow with the dimension $p$, e.g., $m = p$.

Each method returns an estimate for the mixing matrix.
To jointly measure the uncertainty associated with both pre-whitening observations and estimating non-Gaussian components,
we introduce an error measure to evaluate the error between $\widehat{W}_S$ and $W_S^0$ as
\begin{equation*}\label{err}
\min_{Q \in \mathcal{P}^{\pm}_{p \times p}} \frac{1}{\sqrt{pq}}||W_S^0 - \widehat{W}_SQ||^2_\textrm{F}
\end{equation*}
which is similar to the measures in \citet{ilmonen2010new}, \citet{risk2016linear}, and \citet{miettinen2017blind}.
The infimum above is taken such that the measure is invariant to the sign and order of components
with respect to the ambiguities associated with the LNGCA model,
and the optimal $Q$ is solved by the Hungarian method \citep{papadimitriou1982combinatorial}.

We compare the max-min estimator to the max estimator
with various distributions, dimensions of components, and discrepancy functions
in Experiment \ref{exp1} and \ref{exp2} below.

\begin{figure}[ht]
\begin{center}
\centerline{\includegraphics[width=\columnwidth]{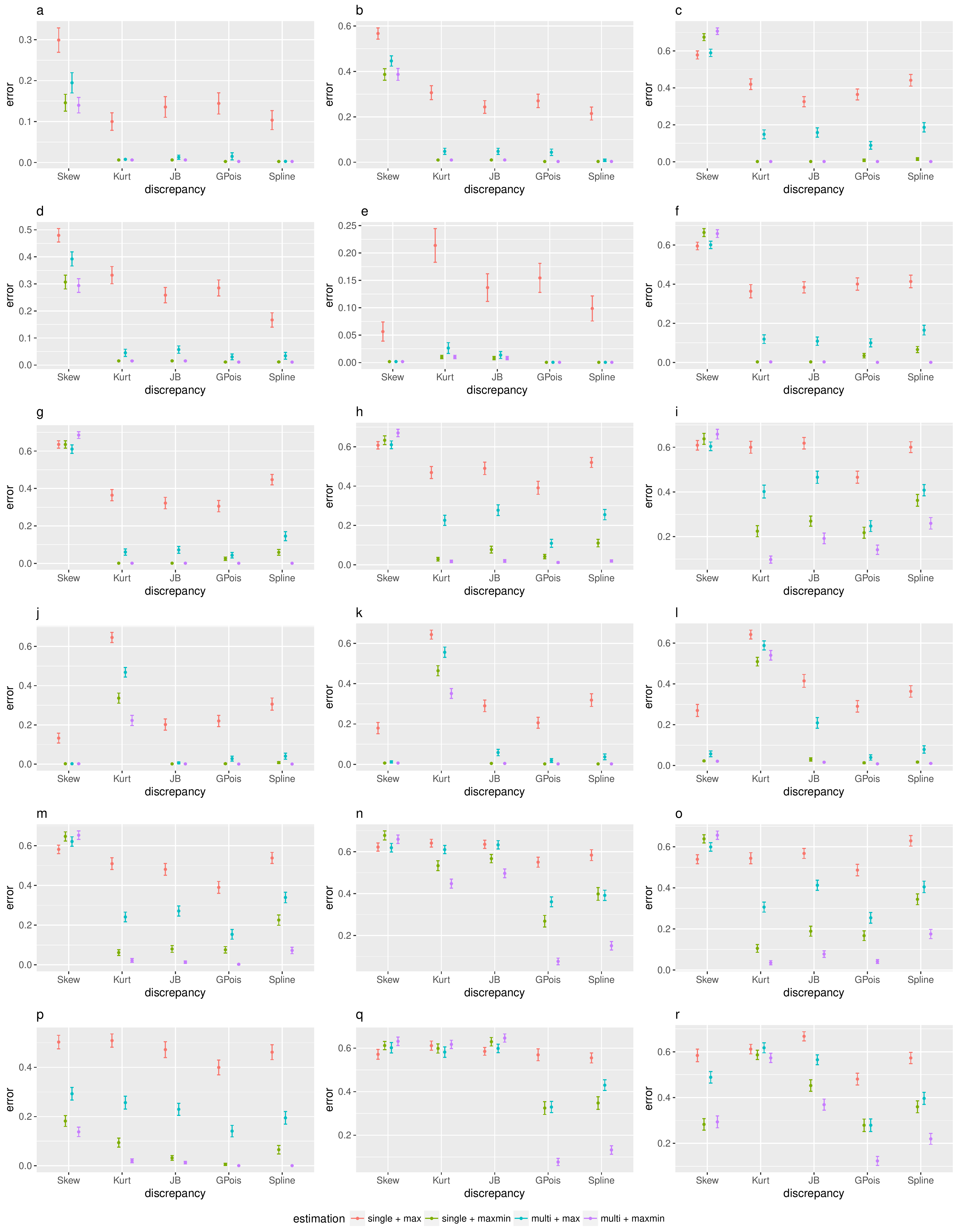}}
\caption{Error measures of both max estimator and max-min estimator with $q = 2$, $p = 4$, $n = 1000$,
100 trials, and both multiple initial points ($m = 4$) and a single initial point ($m = 1$) in Experiment \ref{exp1}.}
\label{fig2}
\end{center}
\end{figure}

\begin{experiment}[Different distributions of components]\label{exp1}
We sample $\mathbf{S}$ from one of the 18 distributions with $q = 2$,
$p = 4$, and $n = 1000$. See Figure \ref{fig2} for the error measures of 100 trials,
with both multiple initial points ($m = 4$) and a single initial point ($m = 1$).

For both multiple initial points and a single initial point, the error measure of the max-min estimator is much lower than that of the max estimator
for most distributions and discrepancy functions.
Therefore, the max-min estimator improves the performance of estimation over the max estimator,
no matter whether a single initial point or multiple initial points is used in optimization.

For both the max-min estimator and max estimator, the error measure with multiple initial points is much lower than that with a single initial point
for most of the distributions and discrepancy functions,
which illustrates the advantage of using multiple initial points over a single initial point.
Moreover, the max-min estimator and multiple initial points turns out to be a powerful combination,
since the error measure of the max estimator with multiple initial points can be further reduced when
replacing the max estimator with the max-min estimator.

The error measure of JB is much lower than that of Skew and Kurt
for most of the distributions, which justifies the joint use of moments.
In addition, GPois is equal and often better than other discrepancy functions
for all the distributions, especially with multiple initial points.
\end{experiment}

\begin{figure}[ht]
\begin{center}
\centerline{\includegraphics[width=\columnwidth]{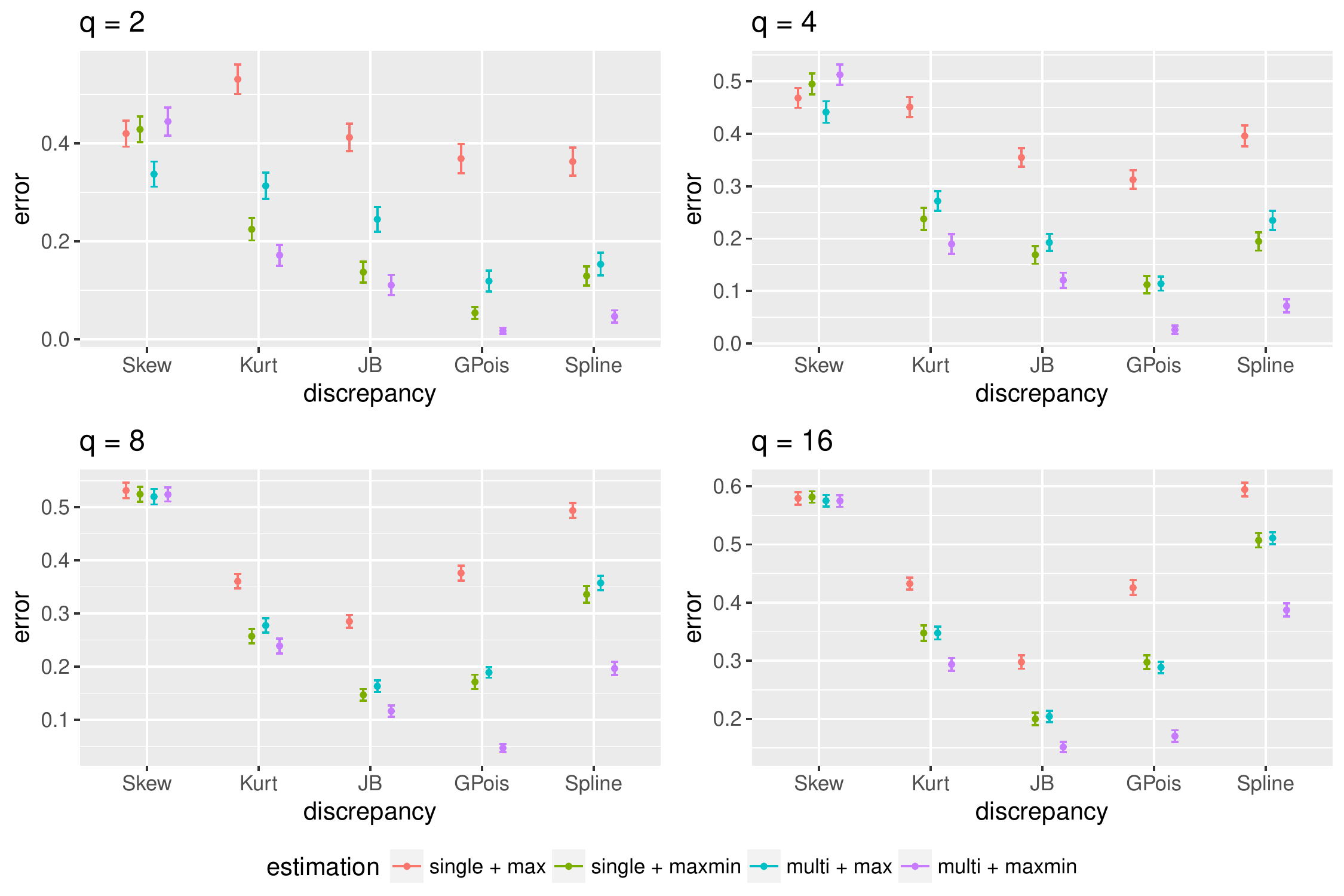}}
\caption{Error measures of both max estimator and max-min estimator with $p = 2q$, $n = 500q$,
100 trials, and both multiple initial points ($m = p$) and a single initial point ($m = 1$) in Experiment \ref{exp2}.}
\label{fig5}
\end{center}
\end{figure}

\begin{experiment}[Different dimensions of components]\label{exp2}
We sample $\mathbf{S}$ from $q$ randomly selected distributions of the 18 distributions, with $q \in \{2, 4, 8, 16\}$,
$p = 2q$, $n = 500q$. See Figure \ref{fig5} for the error measures of 100 trials,
with both multiple initial points ($m = p$) and a single initial point ($m = 1$).




As in the previous experiment, the max-min estimator improves the performance of estimation over the max estimator,
where the error measure with multiple initial points is much lower than that with a single initial point for most cases.
In addition, GPois performs the best for $q = 2, 4, 8$, and JB and GPois perform similarly for $q = 16$
with the max-min estimator and multiple initial points.
\end{experiment}

Since GPois turns out to be more robust to different distributions than Spline in the simulations,
and it shares the same idea with Spline, we omit the results of Spline in the following simulation experiments and data examples.

We compare the current method to the cumulative method for selecting $q$
with various sample sizes of components, and discrepancy functions using the max-min estimator
in Experiment \ref{exp3} below.

\begin{experiment}[Selecting $q$ with varying $n$]\label{exp3}
We sample $\mathbf{S}$ from $q$ randomly selected distributions of the 18 distributions, with $q = 2$, $p = 4$,
$n \in \{2000, 4000, 8000\}$, $B = 200$.
See Table \ref{tab1} and \ref{tab2} for the empirical size and power of 100 trials,
with significance level $\alpha = 5\%$, and both multiple initial points ($m = 4$) and a single initial point ($m = 1$).

For both multiple initial points and a single initial point, the empirical power of the current method is much higher than that of the cumulative method,
while both methods have empirical size around 5\% or even lower,
for all the sample sizes and discrepancy functions.
Hence, the current method outperforms the cumulative method in testing,
no matter whether a single initial point or multiple initial points is used in optimization.

For both the current method and cumulative method,
the empirical size and power with multiple initial points are similar to those with a single initial point,
for all the sample sizes and discrepancy functions,
which implies no remarkable effect in testing from using multiple initial points or a single initial point in estimation.
This suggests that the estimate of the rank of the subspace is less sensitive to initialization than estimates of the individual components.

The empirical power of JB is much higher than that of Skew and Kurt,
for all the sample sizes, which justifies the joint use of moments.
In addition, GPois outperforms the other discrepancy functions,
for all the sample sizes.
\end{experiment}




\subsection{Image Data}\label{data1}

Fulfilling a task of unmixing vectorized images similar to \citet{virta2016projection},
we consider the three gray-scale images from the test images of Computer Vision Group
at University of Granada, depicting a cameraman, a clock, and a leopard respectively.
Each image is represented by a $256 \times 256$ matrix,
where each element indicates the intensity value of a pixel.
Three noise images of the same size are simulated with independent standard Gaussian pixels.
We standardize the six images such that the intensity values across all the pixels in each image have mean zero and unit variance.
Then we vectorize each image into a vector of length $256^2$,
and combine the vectors from all six images as a $256^2 \times 6$ matrix $\mathbf{X}$,
i.e., $p = 6$, $n = 256^2$.
Thus, each row of $\mathbf{X}$ contains the intensity values of a single pixel across all images,
and each column of $\mathbf{X}$ contains the intensity values of a single image.

Then we simulate a mixing matrix $A \in \mathbb{R}^{p\times p}$
using \texttt{mixmat} in the R package \texttt{ProDenICA} \citep{hastie2010prodenica},
and mix the six images to obtain the observations $\mathbf{Y} = \mathbf{X}A^T$,
which are centered by their sample mean, then pre-whitened by their sample covariance to get uncorrelated observations
$\mathbf{Z} = \mathbf{Y}\widehat{H}^T$.
We aim to infer the number of true images, and then estimate the intensity values in them.

First, we run the sequential test to infer the number of true images $q$ with $B = 200$.
See Table \ref{tab3} for the p-values corresponding to each $k$ with a single initial point ($m = 1$).
Both the current method and cumulative method correctly select $q = 3$ with significance level $\alpha = 5\%$,
for all the discrepancy functions.

\begin{figure}[ht]
\begin{center}
\centerline{\includegraphics[width=0.6\columnwidth]{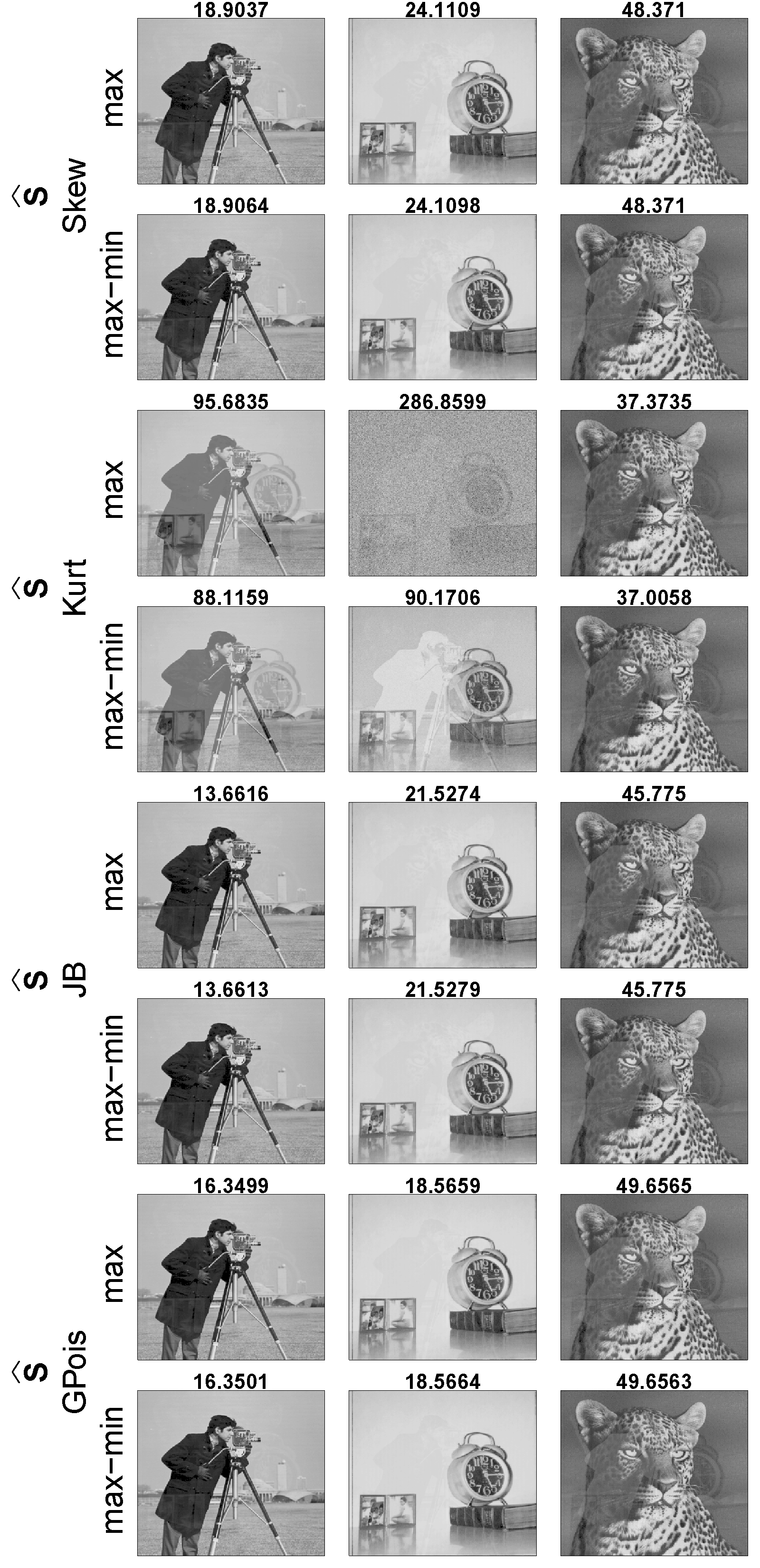}}
\caption{Recovered images of both max estimator and max-min estimator with $q = 3$, $p = 6$, $n = 256^2$,
and multiple initial points ($m = 3$) in estimation for the image data.
Each value on title is the Euclidean norm of the vectorized error image corresponding to the recovered image.
We apply a signed permutation to the images and modify the gray scales for illustration purpose.}
\label{fig6}
\end{center}
\end{figure}

\begin{figure}[ht]
\begin{center}
\centerline{\includegraphics[width=0.6\columnwidth]{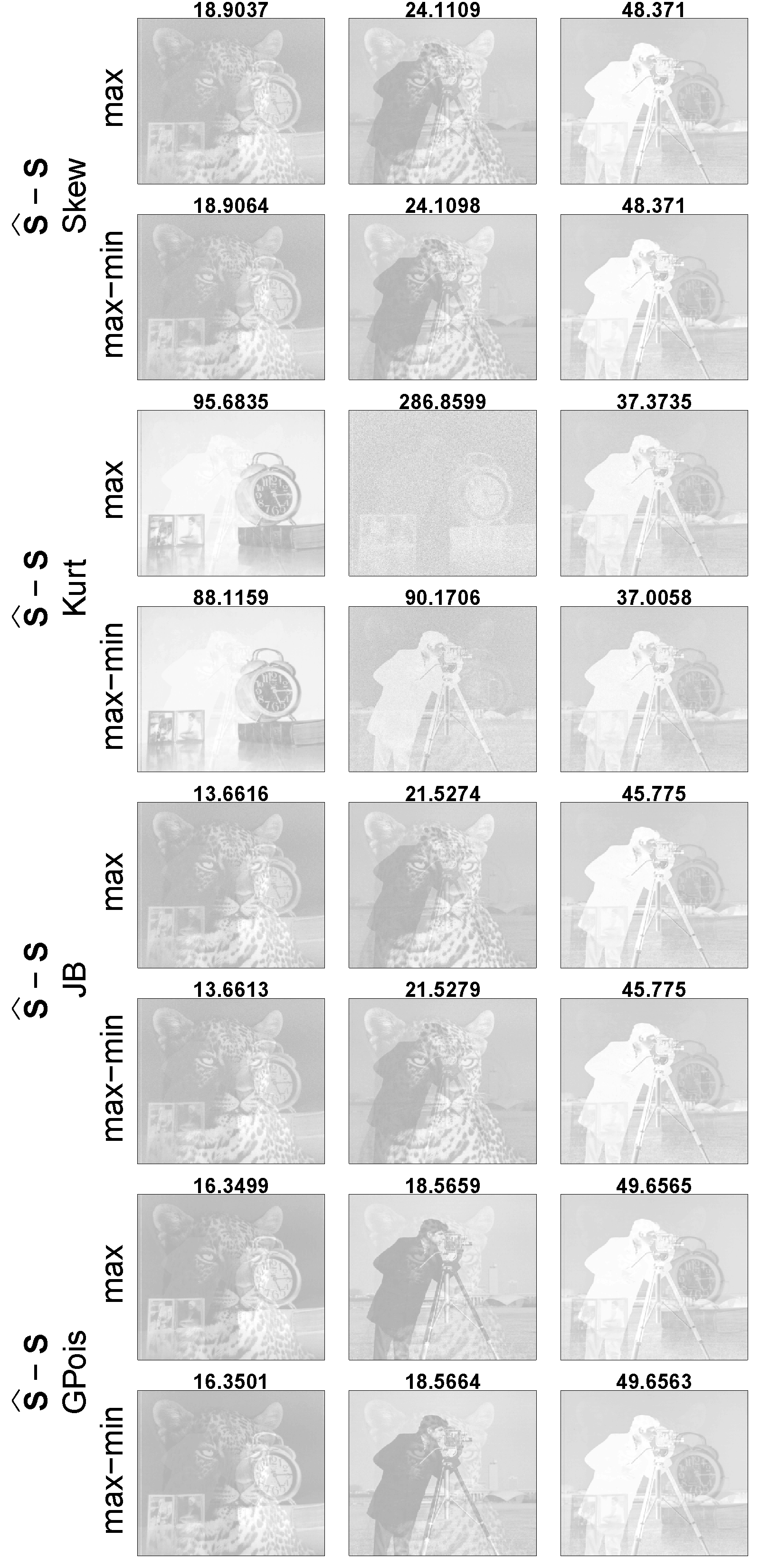}}
\caption{Error images of both max estimator and max-min estimator with $q = 3$, $p = 6$, $n = 256^2$,
and multiple initial points ($m = 3$) in estimation for the image data.
Each value on title is the Euclidean norm of the vectorized error image.
We apply a signed permutation to the images and modify the gray scales for illustration purpose.}
\label{fig7}
\end{center}
\end{figure}

Second, we estimate the intensity values $\widehat{\mathbf{S}}$
with $q = 3$ and multiple initial points ($m = 3$).
See Figures \ref{fig6} and \ref{fig7} for the recovered images $\widehat{\mathbf{S}}$
and error images $\widehat{\mathbf{S}} - \mathbf{S}$,
where the Euclidean norm of vectorized error images is used to evaluate the accuracy of estimation.
The max-min estimator outperforms the max estimator for Kurt,
as the max-min estimator recovers the second image,
while the second image recovered by the max estimator is masked by noise,
and also the max-min estimator has much lower error than the max-min estimator in term of the first image recovered,
which illustrates the advantage of the max-min estimator over the max estimator,
especially when the max estimator does not perform well.
For the other discrepancy functions, both the max-min estimator and max estimator nicely recover the true images.
The estimation of JB is more accurate than that of Skew and Kurt,
as its recovered images are mixed with less noise, indicated by both the estimated images and error images.
In addition, JB and GPois have similar performance, as JB achieves the lowest error on the first image
while GPois achieves the lowest error on the second image.

\section{EEG Data}\label{data}


There are 24 subjects in the EEG data from the Human Ecology Department at Cornell University,
where each subject receives 20 trials.
In each trial, 128 EEG channels (3 unused) were collected with 1024 sample points for a few seconds.
We study the first trial of the first subject. The data of interest is represented by a $125 \times 1024$ matrix,
i.e., $p = 125$, $n = 1024$.
Here, we estimate the number of non-Gaussian signals and examine their time series.
Since the max-min estimator and the current method with GPois perform the best in estimation and testing
of the simulations, we only use the max-min estimator and the current method with GPois in this application.

\begin{figure}[ht]
\begin{center}
\centerline{\includegraphics[width=\columnwidth]{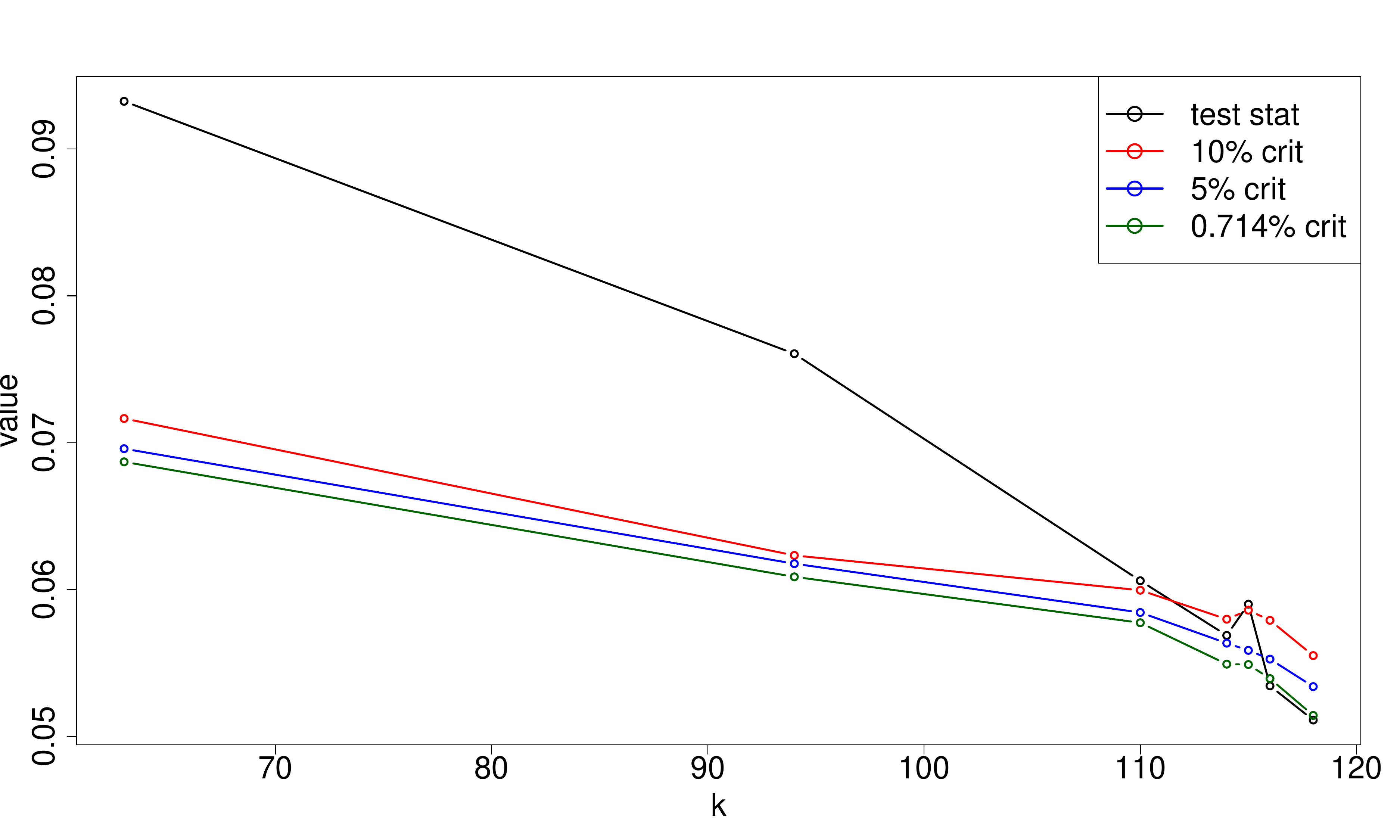}}
\caption{Test statistics and critical values of current method for testing $k$ from binary search
with $p = 125$, $n = 1024$, $B = 200$, and a single initial point ($m = 1$) in testing for the EEG data.}
\label{fig18}
\end{center}
\end{figure}

First, we conduct the sequential test to estimate the number of non-Gaussian signals $q$ with $B = 200$.
Using the binary search for $p = 125$, we expect to have at most $\lceil\log_2 125\rceil = 7$ tests. Hence,
we correct the significance level $\alpha$ to 0.714\% from the original level 5\%.
See Figure \ref{fig18} for the test statistic values (empirical discrepancy) and
critical values at significance level $\alpha \in \{0.714\%, 5\%, 10\%\}$ (i.e., 99.286\%, 95\%, and 90\% quantiles of $\hat{D}(\mathbf{X}_{(k)}^{(b)})$)
corresponding to $k \in \{63, 94, 110, 118, 114, 116, 115\}$ chosen from the binary search with a single initial point ($m = 1$).
The current method rejects the null hypothesis that there are exactly 114 components (p-value $<$ corrected $\alpha$) and fails to reject the null hypothesis that there are exactly 115 non-Gaussian components (p-value $>$ corrected $\alpha$), thus selecting $q = 115$.

\begin{figure}[ht]
\begin{center}
\centerline{\includegraphics[width=\columnwidth]{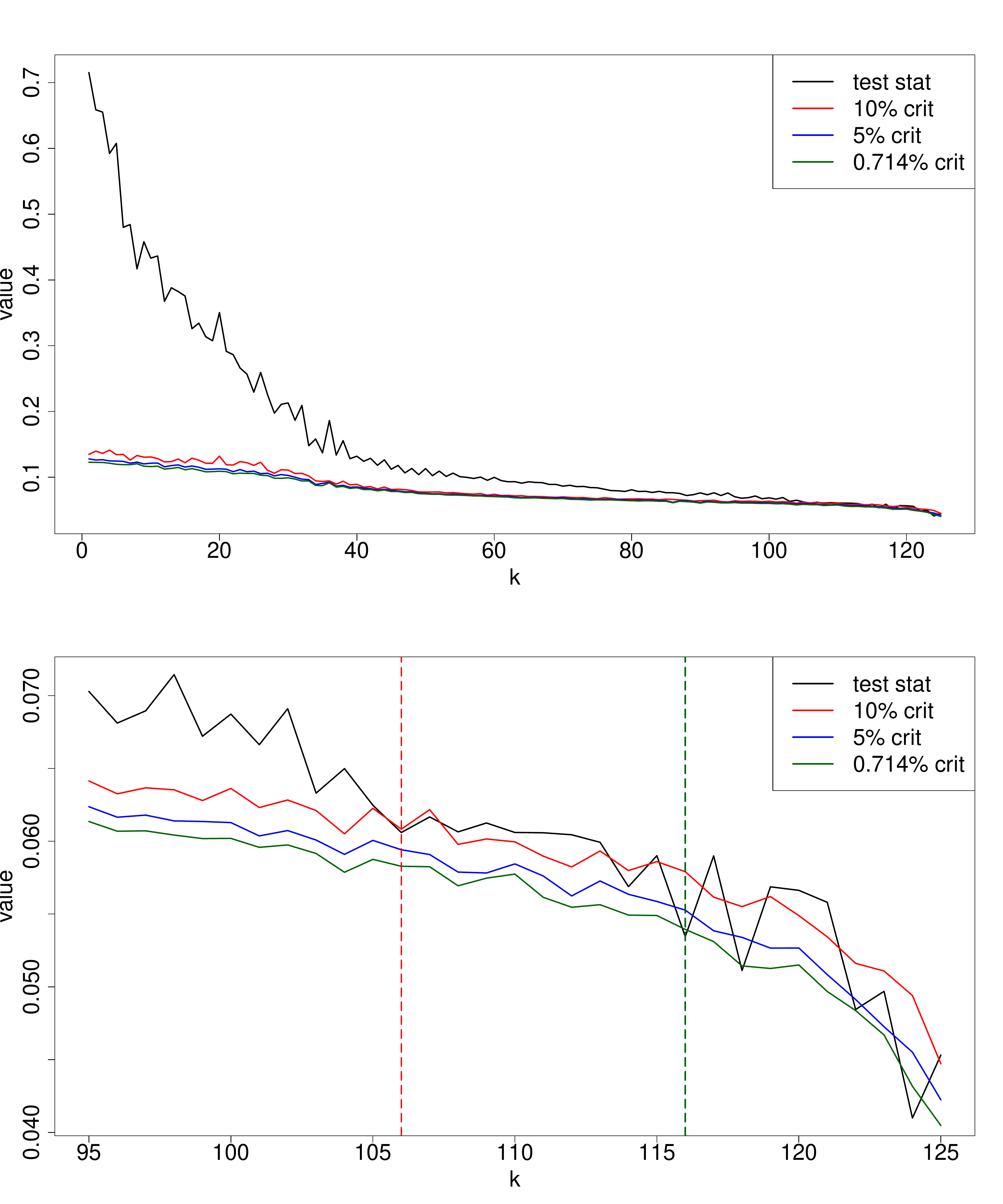}}
\caption{Test statistics and critical values of current method for testing all $k$
with $p = 125$, $n = 1024$, $B = 200$, and a single initial point ($m = 1$) in testing for the EEG data.}
\label{fig19}
\end{center}
\end{figure}

We also iterate all $k = 1, \dots, p$ and provide the complete testing results for reference.
See Figure \ref{fig19} for the test statistic values and
critical values at significance level $\alpha \in \{0.714\%, 5\%, 10\%\}$
corresponding to each $k$ with a single initial point ($m = 1$).
The dashed lines pinpoint where test statistic values meet with critical values,
indicating that this component is assumed to be Gaussian because we cannot reject the null hypothesis.

\begin{figure}[ht]
\begin{center}
\centerline{\includegraphics[width=0.97\columnwidth]{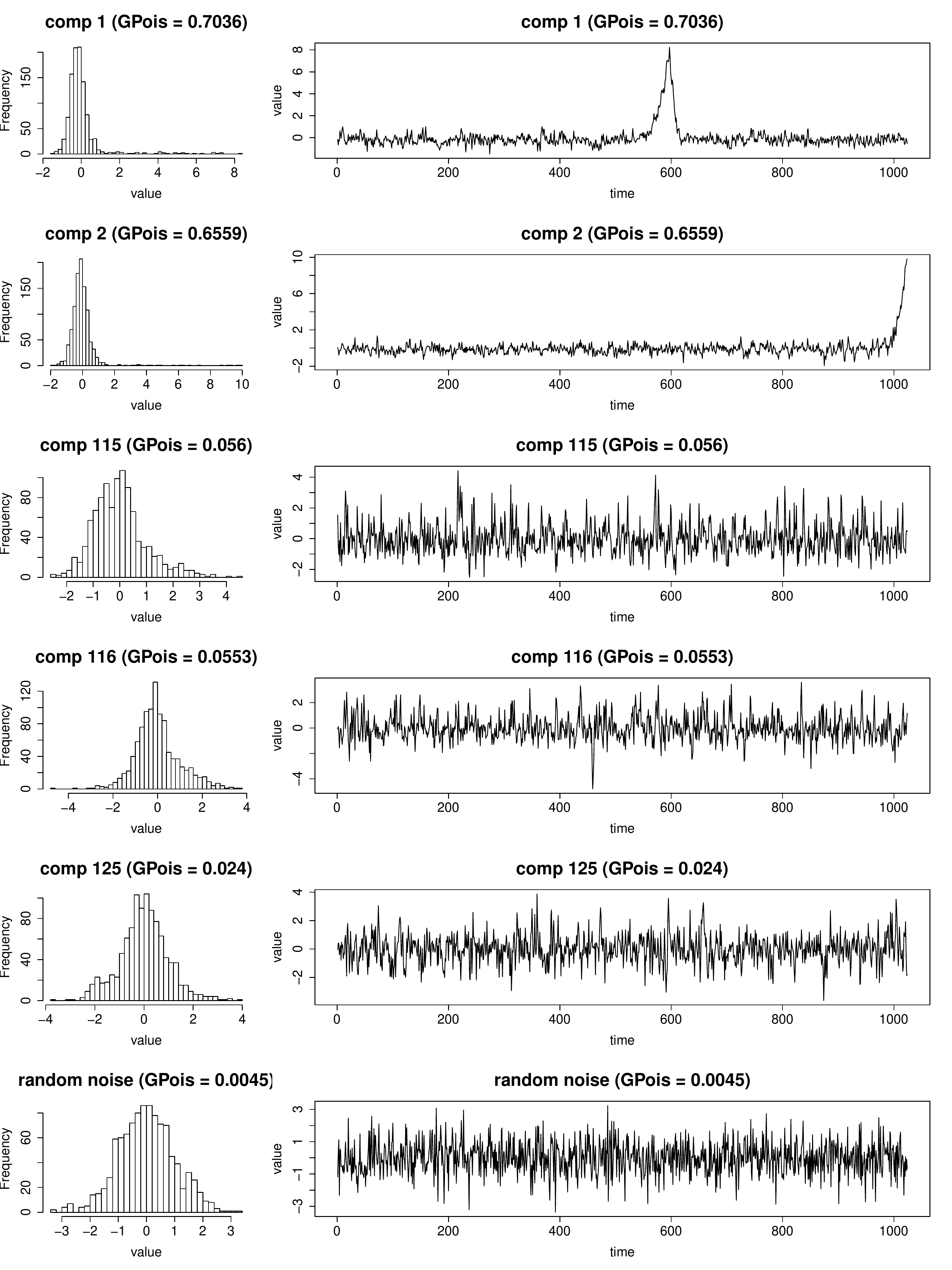}}
\caption{Estimated signals of max-min estimator
with $q = 115$, $p = 125$, $n = 1024$, and multiple initial points ($m = 100$) in estimation for the EEG data.}
\label{fig20}
\end{center}
\end{figure}

Second, we estimate the true signals $\widehat{\mathbf{S}}$
with $q = 115$ and multiple initial points ($m = 100$).
See Figure \ref{fig20} for the estimated signals $\widehat{\mathbf{S}}$.
The max-min estimator successfully extracts meaningful first and second components,
which may be artifacts related to eyeblinks in the middle and at the end of the trial.
The 115th and 116th components are likely to be Gaussian, as they are on the boundary of the p-value = 0.714\%.
The 125th (last) component is fairly close to Gaussian, compared to the Gaussian noise we randomly generate
with the same sample size as a reference distribution.

\section{Conclusion}\label{con}

In this paper, we study the LNGCA model as a generalization of the ICA model, which can have any number of
non-Gaussian components and Gaussian components, given that all components are mutually independent.
Our contributions are the following:

(1) 
We propose a new max-min estimator, maximizing the discrepancy of each non-Gaussian component
from Gaussianity and minimizing the discrepancy of each Gaussian component from Gaussianity simultaneously.
On the contrary, the existing max estimator
only maximizes the discrepancy of each non-Gaussian component from Gaussianity,
which has been used in the ICA model \citep{hastie2003independent} and the LNGCA model \citep{risk2016linear}.
Our approach may seem unintuitive because the individual Gaussian components are not identifiable. However, the Gaussian subspace is identifiable,
and joint estimation of the non-Gaussian components and Gaussian components balances the non-Gaussian subspace
with the Gaussian subspace. This helps shape the non-Gaussian subspace,
and thus improves the accuracy of estimating the non-Gaussian components.

(2) In practice, we need to choose the number of non-Gaussian components.
We introduce a sequence of statistical tests based on generating Gaussian components and ordering estimated components by empirical discrepancy,
which is computationally efficient with a binary search to reduce the actual number of tests.
Two methods with different test statistics are proposed, where the current method considers the discrepancy value of the component under investigation,
while the cumulative method considers the total discrepancy value of all the components
from the first one up to the one under investigation.
Although our test shares some characteristics with that of \citet{nordhausen2017asymptotic},
it has less computational burden with no bootstrap needed and is more flexible in choosing the test statistics.

We evaluate the performance of our methods in simulations,
demonstrating that the max-min estimator outperforms the max estimator given the number of non-Gaussian components
for different discrepancy functions, dimensions, and distributions of the components,
no matter whether a single initial point or multiple initial points is used in optimization.
When the number of non-Gaussian components is unknown, our statistical test successfully
finds the correct number with different discrepancy functions, and sample sizes,
where the current method is more powerful than the cumulative method.

In the task of recovering true images from mixed image data,
our test determines the correct number of true images, and we illustrate the advantage
of the max-min estimator over the max estimator through some discrepancy functions.
Specifically, the max-min estimator nicely recovers the images while the max estimator fails
using the same discrepancy function, and the estimation error of the max-min estimator is equal and sometimes lower than of the max estimator.

In the task of exploring EEG data, our test finds a large number of non-Gaussian signals,
and it extracts two components as the first two non-Gaussian components that may correspond to eye-blink artifacts.
The distributions of estimated signals tend to become more Gaussian as their empirical discrepancy values decrease.
There are a large number of non-Gaussian components in this data set.
In data applications, applying a preliminary data reduction step using principal component analysis (PCA) would likely remove non-Gaussian signals. This underscores the importance of a flexible estimation and testing procedure.

There can be two directions for the future research. One is to look for a better way to address the multiple testing issue in searching a suitable $q$.
Another one is to better understand the improvements with the max-min estimator from a theoretical perspective.
Our intuition is that the contributions of the non-Gaussian components to the asymptotic variances would equal zero.
Therefore, it would be great to gain additional insight into the statistical versus computational advantages of the max-min estimator.

\singlespacing


\bibliographystyle{abbrvnat}
\bibliography{Paper}

\doublespacing

\begin{table}[ht]
\caption{p-values of both current method and cumulative method with $q = 3$, $p = 6$, $n = 256^2$, $B = 200$,
$\alpha = 5\%$, and a single initial point ($m = 1$) in testing for the image data.}
\label{tab3}
\begin{center}
\begin{tabular}{|c|c|c|c|c|c|c|c|}
\hline
Discrepancy & Method & $k = 1$ & $k = 2$ & $k = 3$ & $k = 4$ & $k = 5$ & $k = 6$ \\ \hline
        \multirow{2}{*}{Skew}    & current            & 0.000 & 0.000 & 0.000 & 0.105 & 0.895 & 0.945  \\
                & cumulative                          & 0.000 & 0.000 & 0.000 & 0.815 & 0.970 & 0.975  \\ \hline

        \multirow{2}{*}{Kurt}    & current            & 0.000 & 0.000 & 0.000 & 0.875 & 0.725 & 0.465  \\
                & cumulative                          & 0.000 & 0.000 & 0.055 & 1.000 & 1.000 & 1.000  \\ \hline

        \multirow{2}{*}{JB}      & current            & 0.000 & 0.000 & 0.000 & 0.965 & 0.795 & 0.455  \\
                & cumulative                          & 0.000 & 0.000 & 0.000 & 1.000 & 1.000 & 0.990  \\ \hline

        \multirow{2}{*}{GPois}   & current            & 0.000 & 0.000 & 0.000 & 0.350 & 0.960 & 0.760  \\
                & cumulative                          & 0.000 & 0.000 & 0.000 & 0.625 & 0.715 & 0.675  \\


\hline
\end{tabular}
\end{center}
\end{table}

%

\begin{table}[ht]
\caption{Empirical size and power of both current method and cumulative method with $q = 2$, $p = 4$, $B = 200$,
100 trials, $\alpha = 5\%$, and a single initial point in Experiment \ref{exp3}.}

\label{tab1}
\begin{center}
\begin{tabular}{|c|c|c|c|c|c|c|}
\hline
\multirow{2}{*}{$n$} & \multirow{2}{*}{Discrepancy} & \multirow{2}{*}{Method} & \multicolumn{2}{c|}{power} & \multicolumn{2}{c|}{size} \\ \cline{4-7}
 &  &  & $k = 1$ & $k = 2$ & $k = 3$ & $k = 4$ \\ \hline
\multirow{8}{*}{2000}
    &    \multirow{2}{*}{Skew}    & current            & 0.67 & 0.24 & 0.04 & 0.01 \\
    &            & cumulative                          & 0.67 & 0.13 & 0.00 & 0.00 \\ \cline{2-7}

    &    \multirow{2}{*}{Kurt}    & current            & 0.84 & 0.41 & 0.00 & 0.00 \\
    &            & cumulative                          & 0.84 & 0.18 & 0.00 & 0.00 \\ \cline{2-7}

    &    \multirow{2}{*}{JB}      & current            & 0.92 & 0.60 & 0.00 & 0.00 \\
    &            & cumulative                          & 0.92 & 0.30 & 0.00 & 0.00 \\ \cline{2-7}

    &    \multirow{2}{*}{GPois}   & current            & 1.00 & 0.95 & 0.06 & 0.01 \\
    &            & cumulative                          & 1.00 & 0.94 & 0.00 & 0.00 \\ \hline


\multirow{8}{*}{4000}
    &    \multirow{2}{*}{Skew}    & current            & 0.67 & 0.18 & 0.02 & 0.00 \\
    &            & cumulative                          & 0.67 & 0.13 & 0.00 & 0.00 \\ \cline{2-7}

    &    \multirow{2}{*}{Kurt}    & current            & 0.96 & 0.54 & 0.00 & 0.00 \\
    &            & cumulative                          & 0.96 & 0.28 & 0.00 & 0.00 \\ \cline{2-7}

    &    \multirow{2}{*}{JB}      & current            & 0.99 & 0.78 & 0.00 & 0.00 \\
    &            & cumulative                          & 0.99 & 0.46 & 0.00 & 0.00 \\ \cline{2-7}

    &    \multirow{2}{*}{GPois}   & current            & 1.00 & 0.99 & 0.05 & 0.03 \\
    &            & cumulative                          & 1.00 & 0.99 & 0.00 & 0.00 \\ \hline


\multirow{8}{*}{8000}
    &    \multirow{2}{*}{Skew}    & current            & 0.72 & 0.20 & 0.03 & 0.01 \\
    &            & cumulative                          & 0.72 & 0.18 & 0.00 & 0.00 \\ \cline{2-7}

    &    \multirow{2}{*}{Kurt}    & current            & 0.98 & 0.73 & 0.00 & 0.00 \\
    &            & cumulative                          & 0.98 & 0.46 & 0.00 & 0.00 \\ \cline{2-7}

    &    \multirow{2}{*}{JB}      & current            & 0.99 & 0.90 & 0.00 & 0.00 \\
    &            & cumulative                          & 0.99 & 0.66 & 0.00 & 0.00 \\ \cline{2-7}

    &    \multirow{2}{*}{GPois}   & current            & 1.00 & 1.00 & 0.03 & 0.00 \\
    &            & cumulative                          & 1.00 & 1.00 & 0.00 & 0.00 \\


\hline
\end{tabular}
\end{center}
\end{table}

\begin{table}[ht]
\caption{Empirical size and power of both current method and cumulative method with $q = 2$, $p = 4$, $B = 200$,
100 trials, $\alpha = 5\%$, and multiple initial points in Experiment \ref{exp3}.}
\label{tab2}
\begin{center}
\begin{tabular}{|c|c|c|c|c|c|c|}
\hline
\multirow{2}{*}{$n$} & \multirow{2}{*}{Discrepancy} & \multirow{2}{*}{Method} & \multicolumn{2}{c|}{power} & \multicolumn{2}{c|}{size} \\ \cline{4-7}
 &  &  & $k = 1$ & $k = 2$ & $k = 3$ & $k = 4$ \\ \hline
\multirow{8}{*}{2000}
    &    \multirow{2}{*}{Skew}    & current            & 0.66 & 0.24 & 0.04 & 0.00 \\
    &            & cumulative                          & 0.66 & 0.13 & 0.00 & 0.00 \\ \cline{2-7}

    &    \multirow{2}{*}{Kurt}    & current            & 0.86 & 0.41 & 0.00 & 0.00 \\
    &            & cumulative                          & 0.86 & 0.19 & 0.00 & 0.00 \\ \cline{2-7}

    &    \multirow{2}{*}{JB}      & current            & 0.94 & 0.61 & 0.00 & 0.00 \\
    &            & cumulative                          & 0.94 & 0.30 & 0.00 & 0.00 \\ \cline{2-7}

    &    \multirow{2}{*}{GPois}   & current            & 1.00 & 0.91 & 0.07 & 0.00 \\
    &            & cumulative                          & 1.00 & 0.89 & 0.00 & 0.00 \\ \hline


\multirow{8}{*}{4000}
    &    \multirow{2}{*}{Skew}    & current            & 0.66 & 0.19 & 0.01 & 0.00 \\
    &            & cumulative                          & 0.66 & 0.13 & 0.00 & 0.00 \\ \cline{2-7}

    &    \multirow{2}{*}{Kurt}    & current            & 0.96 & 0.54 & 0.00 & 0.00 \\
    &            & cumulative                          & 0.96 & 0.28 & 0.00 & 0.00 \\ \cline{2-7}

    &    \multirow{2}{*}{JB}      & current            & 0.99 & 0.79 & 0.00 & 0.00 \\
    &            & cumulative                          & 0.99 & 0.47 & 0.00 & 0.00 \\ \cline{2-7}

    &    \multirow{2}{*}{GPois}   & current            & 1.00 & 0.94 & 0.06 & 0.03 \\
    &            & cumulative                          & 1.00 & 0.94 & 0.00 & 0.00 \\ \hline


\multirow{8}{*}{8000}
    &    \multirow{2}{*}{Skew}    & current            & 0.72 & 0.20 & 0.03 & 0.01 \\
    &            & cumulative                          & 0.72 & 0.18 & 0.00 & 0.00 \\ \cline{2-7}

    &    \multirow{2}{*}{Kurt}    & current            & 0.97 & 0.73 & 0.00 & 0.00 \\
    &            & cumulative                          & 0.97 & 0.47 & 0.00 & 0.00 \\ \cline{2-7}

    &    \multirow{2}{*}{JB}      & current            & 0.99 & 0.90 & 0.00 & 0.00 \\
    &            & cumulative                          & 0.99 & 0.66 & 0.00 & 0.00 \\ \cline{2-7}

    &    \multirow{2}{*}{GPois}   & current            & 1.00 & 1.00 & 0.03 & 0.00 \\
    &            & cumulative                          & 1.00 & 1.00 & 0.00 & 0.00 \\


\hline
\end{tabular}
\end{center}
\end{table}

\end{document}